# Ferromagnetism in Fe-doped ZnO Nanocrystals: Experimental and Theoretical investigations

Debjani Karmakar[1], S. K. Mandal [2], R. M. Kadam [3], P. L. Paulose[4], A. K. Rajarajan [5], T. K. Nath [2], A. K. Das [2], I. Dasgupta[6] and G. P. Das[7].

[1] Technical Physics & Prototype Engineering Division, Bhabha Atomic Research Center, Mumbai: 400085, India.

[2] Department of  Physics & Meteorology, Indian Institute of Technology,  Kharagpur- 721302, India.

[3] Radiochemistry Division, Bhabha Atomic Research Center, Mumbai: 400085, India.

[4] Tata Institute of Fundamental Research, Mumbai: 400005, India.

[5] Solid state Physics Division, Bhabha Atomic Research Center, Mumbai: 400085, India.

[6] Department of  Physics, Indian Institute of Technology Bombay, Mumbai: 400076, India.

[7] Department of Materials Science, Indian Association for the Cultivation of Science, Kolkata: 700032, India.

*Abstract*

Fe-doped ZnO nanocrystals are successfully synthesized and structurally characterized by using x-ray diffraction and transmission electron microscopy. Magnetization measurements on the same system reveal a ferromagnetic to paramagnetic transition temperature > 450 K with a low-temperature transition from ferromagnetic to spin-glass state due to canting of the disordered surface spins in the nanoparticle system. Local magnetic probes like EPR and Mössbauer indicate the presence of Fe in both valence states $Fe^{2+}$ and $Fe^{3+}$. We argue that the presence of $Fe^{3+}$ is due to the possible hole doping in the system by cation (Zn) vacancies. In a successive ab-initio electronic structure calculation, the effects of defects (*e.g.* O- and Zn-vacancy) on the nature and origin of ferromagnetism are investigated for Fe-doped ZnO system. Electronic structure calculations suggest hole doping (Zn-vacancy) to be more effective to stabilize ferromagnetism in Fe doped ZnO and our results are consistent with the experimental signature of hole doping in the ferromagnetic Fe doped ZnO samples.

# I. Introduction:

Spintronics is currently an active area of research because spin-based multifunctional electronic devices have several advantages over the conventional charge based devices regarding the data-processing speed, non-volatility, higher integration densities *etc* [1]. The impending need to obtain such devices has led to the growing interest in developing and designing new spintronic materials. Starting from the initial works of Ohno *et. al*. on Mn-doped GaAs which is a ferromagnetic semiconductor with $T_c \sim 110K$ [2], there are continuous efforts to obtain such systems with high $T_c$, preferably close to room temperature[3-5], and also to provide a thorough understanding of the origin of ferromagnetism [6-8] in systems like Mn and Cr doped GaAs, InAs, GaN and AlN. Earlier studies on some (II-VI) based doped semiconducting systems like Mn-doped ZnTe and CdTe [9] were not successful and resulted in $T_c$ values only of the order of a few Kelvin.

The possibility of designing a suitable spintronic material having simultaneously the properties of room temperature ferromagnetism and 100% spin polarization or half-metallicity has brightened after the advent of transition metal doped semiconducting oxides like ZnO [10], $SnO_2$ [11] and $TiO_2$ [12]. A wide range of contradicting experimental results, debating on the success [13-18] versus failure [19-23] of obtaining a $T_c$ above room temperature based on such dilute magnetic oxide systems have injected much excitement on the origin of ferromagnetism in these systems. Among all these systems, ZnO belongs to the list of the most suitable candidates for spintronics materials application due to it's abundance and environment- friendly nature and also due to it's potential as a suitable opto-electronic material with wide band gap ($\sim$ 3.3 eV) and high exciton binding energy of 60 meV. Earlier first- principle electronic structure calculations by Sato et. al. [24-25] suggested that **TM** (TM = Ti, V, Cr, Mn, Fe, Co, Ni, Cu) doped ZnO are ferromagnetic provided the TM doping produces carriers forming a partially filled spin-split impurity band. Theoretical calculations also indicate the possibility of designing a ZnO-based room temperature ferromagnet obtained by n-type carrier doping ($Ga_xZn_{1-x}O$)[25] as well by p-type carrier doping ($ZnN_yO_{1-y}$)[26]. Most of the experimental results show that among transition metal dopants, Co-doped ZnO systems are ferromagnetic both in thin film as well as bulk materials [13,16,27]. Also, there have been few reports of ferromagnetism in Mn [28,29], Ni [30] and some other TM [31], such as Cr or V doped ZnO. Ferromagnetism in Mn-doped system is yet to be understood, as Mn in +2 valence state does not dope any carrier to the ZnO system and hence by first principle studies [24], must lead to antiferromagnetic ground state. However, the experimental success on < 4% Mn-doped ZnO thin film [29] has posed a challenge to the theoretical understanding of ferromagnetism for such systems. Further experimental studies reveal the importance of defects such as interstitials, zinc and oxygen vacancies on the magnetic ordering in such systems [14, 32]. Regarding the experimental studies, it can be commented that the presence of long range ferromagnetic ordering, especially the magnetic moment per cation and $T_c$ depends largely on the critical details of sample preparation. Hence, in order to understand the exact ferromagnetic mechanism in such systems, there are several important issues to be resolved. First, the existence of room temperature ferromagnetism must be related to some intrinsic origin and it must not be due to some impurity phases or TM magnetic clusters. There are studies, which advocate the presence of segregated invert-spinel phase in oxide based dilute magnetic nanoparticle systems [33]. This phase segregation apparently supports magnetic network over the dimension of the crystal, resulting in a high temperature ferrimagnetic phase. Some of the earlier studies [34] correlate the existence of some other impurity phases with the observed results. Second, in most of the thin film studies like reference [11], the magnetic moment per TM atom exceeds the spin-only moments for the corresponding pure TM atom. However, the non-reproducibility of some such studies to obtain giant magnetic moment has strengthened the need for a proper explanation of the observed experimental results. Finally, the exact nature of ferromagnetic interaction between the TM atoms must be probed both experimentally and theoretically.

The present theoretical understanding of ferromagnetism in zinc oxide based dilute magnetic oxide systems is far from complete. The currently existing explanations towards the nature and origin of ferromagnetism in dilute magnetic oxide systems can be categorized as follows:

(a) The spin-split donor impurity band model proposed by Coey *et. al* [35], where the longer ferromagnetic exchange can be mediated by polaronic percolation [36,37] of bound magnetic polarons formed by the point defects like oxygen vacancies. The defect induced shallow donor levels gets hybridized with the TM d-band and thus stabilizes ferromagnetic ground state for such materials [35]. This model can explain the giant magnetic moment observed for thin films and also the ferromagnetism observed for samples with extremely diluted concentration of TM impurity, where there is no possibility of nearest neighbour interaction and the ferromagnetic interaction must be mediated by some other agents.

(b) On the other hand, a recent first principle based study [38] suggests that long-range ferromagnetic coupling can be obtained even in the absence of defects. In this study, the explanation towards the observed ferromagnetism is given by the hybridization picture of super- and double-exchange, where the interaction is mediated by the delocalized Zn-s states. They could improve long range ferromagnetic ordering in Co-doped ZnO system by co-doping it with Li, which induces a non-spin polarized s-like state. Li codoping improves the ferromagnetic ordering by bringing the d-band of TM atom to the optimal position to initiate double exchange. So, according to this study, spin-splitted defect bands are not essential for ferromagnetism.

The above discussion suggests that the mechanism behind ferromagnetism in TM-doped ZnO is not yet fully understood. In the present work, we have investigated the nature of magnetism in Fe-doped ZnO system, both experimentally and theoretically. In Fe-doped ZnO system, there are very few evidences of ferromagnetism [20] except with co-doping [39,40,41]. However, some such co-doping cases are proved to be associated with the formation of nonstoichiometric spinel ferrite phases [33]. From some recent studies [42], the magnetic anisotropy of the dopant cation is suggested to be a signature of intrinsic ferromagnetism in dilute magnetic oxide materials. From the studies of Venkatesan et. al.[17], it is evident that next to $Co^{2+}$, $Fe^{2+}$ doped in ZnO is also highly anisotropic in nature [20]. This suggests the possibility of obtaining high-temperature ferromagnetism in Fe-doped ZnO. Hence, magnetism in Fe-doped ZnO seems to be both promising and interesting issue to investigate. Instead of bulk material, we concentrate on the magnetic properties of Fe-doped ZnO in nanocrystalline form, since a clear understanding of finite size effects on the magnetic mechanism of such systems is essential for the development for high-density magnetic storage media with nanosized constituent particles or crystallites. The effective storage density in a particular device is mostly determined by the stability of the stored information in it, which is largely limited by the spontaneous magnetization reversal in the constituent nanoparticles. Earlier studies in the literature on the Co [43] and Mn-doped ZnO nanoparticles[44] were mostly dominated by superparamagnetism, an widely studied finite size effect.

In our studies, we have mainly focussed our investigations in the following categories. We have prepared Fe-doped ZnO nanocrystals by chemical pyrophoric reaction technique and carried out structural charaterization with XRD and TEM, resulting in a clear nanocrystalline phase without any segregated impurity phase. Different magnetic measurements on this sample reveal a $Tc > 450$ K with a well-defined hysteresis loop at 300K. We have tried to understand the obtained ferromagnetism in terms of the core-shell structure of the underlying nanoparticle system where the spins on the shell adjacent to the surface are in a canted disordered state and the shell is in a perfectly ordered ferromagnetic phase. For nanocrystalline systems, due to their high surface to volume ratio, a critical observation of surface microstructure and behaviour of individual atomic (ionic) moments is essential to analyze the underlying magnetic mechanism [45]. We have tried to explain the observed magnetism in terms of a model where surface defects / vacancies mediate the inter-core / inter-particle ferromagnetic exchange. We have also carried out EPR and Mössbauer studies on the same sample to clarify the issues like percentage of aligned moments and valence state of the TM ions.

Next, we have tried to put forward an explanation of the origin of ferromagnetism by analyzing the electronic structure of the studied system. The theoretical investigations of electronic structure are carried out in the framework of the tight-binding linearized Muffin-tin orbital method (TB-LMTO) in the atomic sphere approximation (ASA)[46] within local spin density approximation (LSDA) [47,48]. The effects of point defects such as oxygen vacancies and cation vacancies on the ferromagnetic properties of such systems are also investigated.

The paper is organized as follows. In section II, we describe the preparation and structural characterization of the sample. In the successive sections III, IV and V, we describe respectively magnetization measurements, EPR and Mössbauer studies. The possible explanations towards the observed results are provided in section VI. Section VII is devoted to the theoretical calculations and discussion of the results. The last section summarizes all the results with a conclusion.

## II. Sample preparation and Structural characterization

Most of the bulk studies on TM doped ZnO polycrystalline materials reported in the literature [19, 20, 21, 33, 34, 39, 40] were carried out on materials prepared using solid-state reaction technique. In the present work, we utilized chemical route for sample preparation, which, being a comparatively low temperature technique, is useful to prepare single-phase nanocrystalline samples.

Nanocrystalline 10 % Fe doped ZnO powders are synthesized by using chemical pyrophoric reaction method. Requisite amount of $Zn(NO_3)_2.6H_2O$ and $Fe(NO_3)_3.9H_2O$ are dissolved in distilled water depending on the percentage of Fe-doping. This solution was heated at 190 $^0$ C with constant stirring. After some time Tri-ethanol amine is added in a ratio 4:1 with the metal ions in order to precipitate the metal ions with the amine. At the same temperature $HNO_3$ is added to dissolve the precipitate and the resulting clear solution was allowed to evaporate with continuous stirring at 190 $^0$ C. After complete dehydration, we got the precursor powder, which after grinding was calcined at $350^0$C. The calcination temperature was optimized by repeated investigations such that the secondary iron oxide phase formation could be avoided. In a previous study [49], the optimization procedure for preparing single-phase nanocrystalline systems reveals that both the percentage of TM-dopant and the calcination temperature are appropriate for avoiding the secondary phase formation. The secondary phase in the present case was identified with $Fe_3O_4$ [49]. Transmission electron microscopy (TEM) reveals that the powder samples are nanocrystalline in nature with a broad size distribution from 2 nm to 30 nm with an average size of 7 nm. The structural charaterizations have been carried out by using a standard x-ray diffractometer (Phiips, PW-1729) with monochromatic Cu-K$\alpha$ radiation and by high-resolution transmission electron microscope (JEOL, JEM-2010, 200 kV).

Powder X-ray diffraction patterns for pure ZnO and 10% Fe doped ZnO are presented for comparison in Fig. 1(a) and (b) respectively. Due to the smaller particle size, the XRD peaks of the doped sample are broadened. The profile fitting of the data was performed by Le-Bail method using GSAS program [50] to find out changes in the unit cell volume for different compositions of Fe. The fitted and the observed data alongwith the difference between them are indicated in the figures 1(a) and (b). To investigate the solubility of Fe in ZnO, we have studied the XRD for four different compositions of $Zn_{1-}$ $_x$Fe$_x$O; with x = 0.0, 0.05, 0.1 and 0.15. Since, in this paper we concentrate on the 10% Fe doped ZnO, the XRD data and the fits are shown for x = 0.0 and 0.1 only. Except x = 0.15, none of the samples in the compositional range have shown any evidence for impurity phases. All the data can be fitted in the hexagonal wurtzite structure. With the increase of Fe content, the x-ray peak widths increase suggesting a decrease in crystalline correlation. By cell parameter refinement using Le-Bail method, we have calculated the evolution of the cell volume as a function of Fe-concentration, plotted in Fig. 2. For the first three concentrations, the cell volume increases almost linearly, whereas for the fourth one (15%), it decreases, indicating that the limiting composition must be less than that. The situation is quite similar to references [20] and [22].

Fig. 3(a)-(d) show the results of TEM characterization. Fig. 3(a) shows the low-magnification TEM micrograph for 10% Fe doped ZnO sample. The micrograph reveals broad particle size-distribution with the diameter of the particle ranging from 2 nm to 30 nm. The high-resolution TEM (HRTEM) micrographs are presented in Fig. 3(b) which shows that all the nanoparticles are single crystalline and free from any major lattice defects. Crystallinity and preferential orientation of each nanoparticles in the sample are confirmed from selected area diffraction patterns (SAD) shown in Fig. 3(c). The SAD pattern obtained by focusing the beam on a few nanoparticles of the sample clearly indicates the single crystalline nature of each nanoparticle. Also, it confirms that the nanocrystals are indeed in the wurtzite phase. The particle diameters are estimated from the low-resolution TEM data using the standard software (Image-J) and the particle diameter histogram is fitted with lognormal distribution function in Fig. 3(d). The average particle size is obtained as 7 nm.

# III. Magnetic measurements:

With a successful synthesis and structural characterization, we proceed with the magnetic measurement for the 10% Fe-doped ZnO nanocrystals. The percentage of doped TM metal is, of course, slightly on the higher side towards the cationic percolation threshold for this material. The average particle size, as obtained from the low-resolution TEM data is around 7 nm, whereas a broad particle size distribution exist within the material. At the present situation, on an average, surface atoms constitute nearly 15% of the total number of atoms, the percentage being more (~ 25 %) for smaller size of particles (1-7 nm). Hence, the influence of surface atoms on the magnetic behaviour is not negligible. We have attempted to understand the observed behaviour in terms of the core-shell structure for each nanoparticle. The surface atoms are different from the core ones, as the total number of exchange interactions will be less for them due to their lower co-ordination (fewer nearest neighbours). Moreover, there can be broken exchange bonds [45] on the surface due to cationic vacancies, oxygen vacancies or due to the presence of organic surfactant molecules. As proposed by Kodama *et. al.* [45], these broken exchange bonds are largely responsible for the surface spin disorder, the percentage of which is more for smaller size of particles. Further, the basic reason for irreversibility in the magnetization is due to magnetocrystalline, magnetoelastic and shape anisotropies. Magnetocrystalline anisotropies for the surface and the core atoms are different due to the easy-axis nature of the anisotropy for the surface, as proposed by Neel [51], in contrast with the easy plane anisotropy for the core. Hence, the magnetic behaviour will be different for the surface and core atoms and will be dominated either by core or by surface depending on the factors like applied field, temperature, percentage of aligned spins etc.

Earlier studies on the Co-doped ZnO nanoparticles prepared by vaporization condensation method [43] and Mn-doped ZnO nanoparticles prepared by co-precipitation technique [44] have observed a ferromagnetic behaviour at low temperature and superparamagnetic behaviour at high temperature. For some of the samples, depending on the preparation preocedure, Curie-Weiss behaviour is observed confirming an antiferromagnetic ground state.

In our experimental observation, we have obtained room temperature ferromagnetism for 10 % Fe doped ZnO nanocrystals. The magnetic measurements were carried out using a Quantum design MPMS SQUID Magnetometer and a Vibrating sample magnetometer (VSM). Temperature dependence of magnetization is investigated for the sample (in the form of a pressed pellet) both in ZFC (zero-field cooled) and FC (field cooled) conditions for four different field values, H = 100 Oe, 1000 Oe, 1500 Oe and 5000 Oe, as depicted in Fig. 4(a)-(d). The outward concave nature of all these magnetization curves confirms the non-mean-field like behaviour for the presently studied system, where the magnetic behaviour is explained in terms of carrier induced effects. The concave shape confirms the low carrier density and the localized nature of the carrier as well [52]. From Fig. 4(a), it is evident that the transition temperature is > 320 K, the maximum permissible instrumental limit for us. With the gradual increase of the applied field value, the meeting point of ZFC and FC curves shifts towards lower temperature and at a sufficiently high field, there will be absolutely no irreversibility. Figure 4(a) indicates that the system will be weakly ferromagnetic even at room temperature, unlike the other two recent studies [43,44], where the superparamagnetism is the predominant magnetic mechanism at high temperature. The M-vs-T curves obtained in the present study has a basic difference in their nature with superparamagnetism dominated curves in reference [43] for Co-doped ZnO nanoparticles as well as reference [53] for highly reduced Co-doped $TiO_{2-\delta}$ thin films, where the FC and ZFC curves diverge substantially at low temperatures. From the first three graphs, 4(a), (b) and (c), the ZFC curves are all having a cusp which with the increase of applied field shifts towards low temperature. For a magnetic system, the cusp in the ZFC curve may appear due to various reasons. It may result due to the nanoscale related superparamagnetic behaviour or due to field-induced cross-over from ferromagnetic (FM) to Spin-Glass (SG) state as observed in mixed valence manganites [54]. To investigate the actual reason of appearance of the cusp, we have measured the relaxation in the thermoremnant magnetization. The sample is field cooled in a field of 1000 Oe upto 5 K, which is below the peak of the cusp temperature corresponding to the field (as can be seen in Fig 4(b)), and then the field is switched off. The magnetization of the sample is measured using a VSM, as a function of time. The magnetization vs. time data is plotted in figure 5(a), which indicates a considerable amount of relaxation within a time period of 7000 seconds. This curve rules out the possibility of cusp behaviour in the ZFC curve due to the superparamagnetic behaviour as for the superparamagnetic systems the relaxation is very fast being within the experimental time scale of hundreds of seconds [55]. Also, the time dependence of thermoremnant magnetization can be fitted well with a stretched exponential function $M(t) = M_0 + M_r \exp[-(t/\tau)^{1-n}]$, where $M_0$ is related to the intrinsic ferromagnetic like component and $M_r$ relates to a glassy component contributing to the observed relaxation effect. The time constant $\tau$ and the parameter $n$ are related to the relaxation rate. Stretched exponential functions are usually used to describe the magnetic relaxation in spin-glasses and cluster spin-glasses [56]. We have also measured the same relaxation at 50 K (not shown) and have not observed any relaxation as it is above the spin glass transition temperature.

The existence of this spin glass state may be explained in terms of the core-shell type model for the spin-structure of the underlying nanoparticle system in the following way. For a particular temperature and applied field, the core spins are aligned in a particular direction. For a given orientation of the core magnetization, there can be multiple metastable states (although, for convenience, only one is shown in Fig. 5(b)) for surface spin configurations, separated by rotational barriers, which can be crossed by increasing thermal energy. For a ZFC sample, at very low-temperature, as a small field is applied (e.g.100 Oe, as seen in Fig. 4(a)), due to canting of the surface spins in a disordered configuration, the system remains in a spin glass state, as depicted in Fig. 5(b), where there is no long range order. With the increase of temperature, the thermal energy helps the disordered surface spins to get themselves oriented (by providing the required energy to cross the barrier) along the resultant core magnetization direction and thus to drive the whole system in a more stable ferromagnetic configuration. The ferromagnetic configuration is mostly dominated by the core magnetization, where the exchange interactions among cores in different nano-particles are mediated by various surface disorders like cation or anion vacancies. With gradual increase of the applied field, more and more cores spins are connected by ferromagnetic exchange and the surface spin effects are less pronounced (as the total number of disordered surface spins are less compared to the ordered core spins) leading to a lowering of the spin-glass to ferromagnetic transition temperature. For a high field value (*e.g.* 5000 Oe, as seen in Fig. 4(d)), the mechanism is totally controlled by the core magnetization and hence there is no irreversibility. The whole situation is explained by the multiple potential-well model, as presented in fig. 5(b), where the force to cross the metastable barrier is provided either by the magnetic field or by temperature.

In Fig. 6(a), for the field values 1000 Oe and 1500 Oe, we have calculated the dc susceptibility $\chi_{dc}$ from the magnetization data and plotted in a Curie-Weiss like plot. It appears from the curves in Fig. 6(a), that above the temperature corresponding to the meeting point of FC and ZFC curves (as can be seen from Fig. 4), both the curves can be fitted with the Curie-Weiss type linear behaviour resulting in high negative Weiss temperatures. This may apparently indicate an antiferromagnetic ground state for the system. For an accurate determination of the exact nature of intrinsic magnetism for the system, we have plotted the inverse differential susceptibility $1 / \chi_{diff}$ (taking the difference of the M-vs-T FC plots for 1500 Oe and 1000 Oe) as a function of temperature in Fig. 6(b), since the differential susceptibility plots can more certainly decide about the exact ground state [22]. From this plot, it is very much obvious that there are two kinds of magnetic phases co-existing in the system. The dotted line in figure corresponds to the paramagnetic Curie behaviour (C/T-type) and the dashed line corresponds to a fit for ferromagnetic Curie-Weiss behaviour (C / (T - θ)) with a positive θ of 59.5 K. This situation implies that some randomized surface spins are always present in the system, giving rise to a paramagnetic contribution superimposed with the core-dominated ferromagnetism. This is indeed supported by the hysteresis loop measurements. The Curie-Weiss fit results into an effective magnetic moment of 0.334 $\mu_B$ from the relation $\mu_{eff} = (3k_BC/N_A)^{-1/2}$, ($N_A$ - Avogadro number), which is much less than the theoretically calculated value. The moment per magnetic cation can be less due to several factors. Since the concentration of Fe is very close to the cationic percolation threshold, nearest neighbour antiferromagnetic interaction (superexchange) between the Fe ions can lower the magnetic moment. Also, the presence of uncoupled $Fe^{3+}$ spins on the surface of the nanoparticles may lower the magnetic moment. The presence of any small randomly distributed cluster (consisting of Fe free spins or due to iron oxides), undetected by XRD can also be the cause for the same. Moreover, due to nanostructured nature of the material, weaker interparticle exchange is also responsible for low magnetic moment.

The M-vs-H hysteresis loops are presented in figure 7(a), (b) and (c). Figure 7(a) shows a comparison of the low and high temperature hysteresis loops. As expected, in low temperature loops (2 K, 5 K, 20 K), as seen in Fig. 7(b), the paramagnetic contribution is high (may be due to disordered $Fe^{3+}$ spins on the nanoparticle surface) and at 300 K, that contribution lowers. As is seen in Fig. 7(a), the sample has a well- defined hysteresis at room-temperature with a coercive field of 94.4 Oe and a remnant magnetization of 0.0788 emu/gm. The corresponding values at 5K are 177.8 Oe and 0.177 emu/gm. The saturation in magnetization at low temperature (2K) is achieved at a very high field, as shown in Fig. 7(c). The magnetic moment per Fe as calculated from the 2K magnetization data is ~ 1.13 $\mu_B$ and at the room temperature that value decreases down to ~ 0.05 $\mu_B$. The obtained hysteresis at room temperature supports the fact that the magnetic behaviour is weakly ferromagnetic and not superparamagnetic. To verify this in a different way, we have also plotted the $M / M_s$ ($M_s$ = saturation magnetization) versus $H / T$ for the $M$-$H$ data at different temperatures and all the curves remain distinct. For superparamagnetic systems, as is well-known, the curves merge into a single one [43,55]. This rules out the possibility of thermal energy induced spontaneous magnetization reversal as happens in the case of an assembly of single domain particles leading to superparamagnetic behaviour. More convincingly, the superparamagnetic limit for Fe-doped systems can be calculated by the expression for the relaxation time $\tau = \tau_0$ exp [$E_A V$ / $k_B T$], where the anisotropy energy density $E_A$ having a value of ($5 \times 10^4$ J / $m^3$ for Fe), $V$ and $k_B$ are the particle volume and Boltzman constant respectively. For superparamagnetic systems, taking $\tau_0 \sim 10^{-9}$ s and the typical measurement time to be $\sim$ 100s, at 5 and 300K, the critical nanopartical diameters for superparamagnetic behaviour are 4 nm and 8 nm respectively. Hence in the presently studied system, at 5 K, as there are very few particles of 4-nm diameter, there are no chance of obtaining superparamagnetism. At 300 K, although some of them are within the critical limit, as there is a wide size distribution, the system behaves as a ferromagnetic one. Moreover, unlike the manganite [56] or ferrite [45] nanoparticles, the irreversibility obtained in this study is not an outcome of only surface spin disorders, an obvious signature of which is very high field irreversibility and shifted hysteresis loops as seen in references [56,45]. None of these signatures are seen in the present study. Hence, the ferromagnetism obtained can be an outcome of some intrinsic magnetic nature of the material studied. In addition to this, although we have obtained room temperature ferromagnetism for our sample, the magnetic moment per Fe ion is much lower than the spin-only calculated limits. To investigate the actual reason for the observed ferromagnetism and low value of magnetic moment per Fe, we have carried two more magnetic studies, *viz.*, the EPR and FMR measurements and also the Mössbauer studies.

## IV. EPR Measurements

To probe the exact electronic configuration and oxidation state of the dopant TM ion more critically and also to understand the ferromagnetic mechanism of the sample at a microscopic level, electron paramagnetic resonance (EPR) measurements were carried out for the present sample. EPR experiments were carried out using a Bruker spectrometer at X-band ($\nu$ = 9.60 GHz) equipped with 100 kHz frequency modulation. Temperature was varied in the range 100 - 450 K using variable temperature accessory Eurotherm B-VT 2000. DPPH was used as a reference for calibration of g factors. The shape and the area of the EPR spectra were analyzed by standard numerical methods.

Electron Paramagnetic Resonance (EPR) is an effective tool to investigate the origin and nature of observed ferromagnetism in a material. The temperature dependent changes in the line position, integrated intensity and line width in the EPR spectra can be used to obtain information about the range of magnetic ordering, spin fluctuations, spin glass behaviour etc. In addition to this, the technique is used to extract informations about the oxidation state of the dopant cation involved in the spin coupling [44, 57-59]. Fig.8 (a and b) shows the EPR spectra of $Zn_{0.9}Fe_{0.1}O$ recorded at room temperature and at 100 K. Both of the EPR spectra of $Zn_{0.9}Fe_{0.1}O$ can be considered as a superposition of two overlapping signals, an intense and broad Gaussian signal ($\Delta H_{pp}$ = 1675 G) with effective g-factor, $g_{eff}$ = 2.126 and another weak and narrow Lorentzian signal with $g_{eff}$ = 2.005. For the sake of convenience, the broad and narrow signals are designated as signal $A$ ($g_{eff}$

= 2.126) and signal *B* ($g_{eff}$ = 2.005) respectively. In the temperature range of the investigations, the effective *g* factor ($g_{eff}$) for the broad signal (signal *A*) was observed to be greater than 2, indicating a clear signature of ferromagnetism in this material. At highest temperature of investigation (450K), the line width of signal *A* decreased to $\Delta H_{pp}$ ~1350 G and $g_{eff}$ = 2.0846. Signal *A* is attributed to ferromagnetic resonance (FMR) arising from exchange interactions between the Fe-ions. It may be mentioned in passing, that such a temperature dependent changes in the line width and line position of the resonance signal would not occur in the paramagnetic state. The observed broadening of the signal and a shift of the center of resonance to the lower fields is due to presence of non- homogenous local magnetic field, which modifies both the resonance field and the line shape of the signal. This effect is greater at lower temperatures [60]. The inset of Fig. 8 depicts a measure of the total number of spins participating in producing signal A and B. It is evident from the figure that with increasing temperature, the number of spins taking part in ferromagnetic interaction (signal A) increases at the cost of decrease in the number of spins participating in paramagnetic signal B.

On lowering the temperature from 450 K to 100 K, the line shape and the resonance field of signal *B* remained practically unchanged, however, an increase in the signal intensity was observed. Whereas signal *A* showed gradual increase in the line width ($\Delta H_{pp}$ = 2000 G at 100 K) with the lowering of the temperature, signal B does not show any such signature and thus indicating it's non-ferromagnetic origin. Fig. 9 depicts the temperature dependence of EPR spectra recorded in the range of 100 – 450 K. With the increase of the temperature, line shape becomes more symmetric and the resonance field increases to higher fields. Due to smaller value of the spin lattice relaxation time,

EPR of $Fe^{2+}$ signal cannot be resolved within the experimental temperature range. The signal B is attributed to the presence of uncoupled $Fe^{3+}$ ion ($^6S_{5/2}$, S= 5/2), the line width and position matching with references [61,62], giving rise to additional paramagnetic contribution. For more critical investigation of the ferromagnetic (signal *A*) and paramagnetic (signal *B*), the peak-to-peak line width ($\Delta H_{pp}$), peak area, intensity and the resonance field or line position ($H_r$) is plotted as a function of temperature for signal A in Fig. 10(a). As expected for the FMR signal, the line width, area and intensity decrease as a function of temperature, whereas the line position increases with temperature. However, below 300 K, as the signal is a combination of two spectra, determination of the line position may become erroneous. Hence, the line position is plotted only for temperatures greater than 300K. It is evident from the EPR spectra, the system remains ferromagnetic upto 450 K. As there is almost no change in the line-width and line position, for signal B, we have plotted only area and intensity for signal B as a function of temperature in Fig. 10(b). Both of these quantities  decrease with the increase of temperature.

To rule out the possibilities of formation of the impurities phases during the synthesis, which may be responsible for the observation of the room temperature ferromagnetism, EPR spectra  were also recorded for high purity $Fe_3O_4$ (identified with the secondary phase) and $Fe_2O_3$ samples at room temperature (figure not shown). The EPR spectra of these compounds were much different than those observed under present investigations.

Since, EPR measurements have shown some signature of presence of doped iron in two valence states $Fe^{2+}$ and $Fe^{3+}$, we wanted to resolve that issue by Mössbauer spectroscopic measurements.

## V. Mössbauer spectroscopy

To probe the local magnetic environment prevailing around the Fe sites and also to determine the oxidation state of Fe in ZnO matrix, Mössbauer spectra was recorded for the sample as a function of temperature.

The $^{57}$ Fe Mössbauer spectra were recorded using a conventional constant acceleration velocity drive and temperature was varied in the range 4 K- 300 K using a continuous flow cryostat. The spectra were recorded in transmission geometry with a $^{57}$Co(Rh) source.

Figure 11 depicts the Mössbauer spectra recorded at three different temperatures *viz.*, at 300 K (room temperature), 12 K and 4 K alongwith the least square fitting. From all these three spectra, presence of uncoupled $Fe^{3+}$ is evident, since all of them indicate the presence of a paramagnetic doublet [62- 65]. The doublet becomes more and more asymmetric with lowering temperature due to the change in the magnetic environment within the sample. The isomer shift for the 300 K, 12K and 4K spectra are 0.56 mm/s, 0.62 mm/s and 0.55 mm/s respectively. The initial increase and subsequent decrease of the isomer shift may happen due to the transition of the sample to the spin glass state with lowering of temperature. This transition due to the freezing of the surface spins changes the local environment of the $Fe^{3+}$ ions and thus affects the s-electron density. More critical observation of the low temperature Mössbauer data (12K and 4K) revealed some very weak signal of sextet hyperfine splitting. Due to the low magnetic moment value for the sample and very small natural abundance of $Fe^{57}$, the intensity of the observed signal is very weak. However, a more detailed fitting of the experimental data for 12 K in Fig. 12 indicates the presence of quadrupole splitting for both $Fe^{3+}$ and $Fe^{2+}$. In the same figure, we have also indicated the position of sextets. The observation of distinguished $Fe^{2+}$ signal is difficult due to it's very short spin-lattice relaxation time. The values for the quadrupole splitting for $Fe^{3+}$ and $Fe^{2+}$, as obtained from Fig. 12 is 0.73 mm/s and 1.4 mm/s respectively [65,66].

Hence both from EPR and Mössbauer measurements, we have obtained an evidence of the existence of both $Fe^{3+}$ and $Fe^{2+}$ within the sample. Our next aim is to qualitatively explore various possibilities to understand the current experimental scenerio.

## VI. Discussion about the presence of $Fe^{3+}$

Usually, if Fe is present in the substitutional site in a defect free ZnO crystal, the valence state of Fe will be 2+. However, both EPR and Mössbauer results confirm the presence of uncoupled $Fe^{3+}$ within the sample, giving rise to the associated paramagnetic behaviour. In this section, we discuss the various possibilities for the presence of iron in both valence states for Fe-doped ZnO system.

The uncoupled $Fe^{3+}$ ions may be due to the presence of nearest neighbour cationic vacancies and thus effectively doping holes in the system. In other words, if cationic vacancies (Zn for the present case) are present in the nearest neighbourhood of $Fe^{2+}$ in the substitutional cationic site, to neutralize the charge imbalance, the valence state of Fe can be converted to 3+. This can happen mostly on the surface of the nanoparticle, where the probability of presence of vacancies is more [67]. The increase in the intensity of the EPR signal $B$ with the lowering of the temperature may be due to freezing of these $Fe^{3+}$ spins. This observation is consistent with the increase in the paramagnetic contribution as seen in the low temperature hysteresis loops. For Mn-doped ZnO nanoparticles, as seen in reference [44], FMR signal was observed at room temperature for nominal 2 at% $Mn^{2+}$: ZnO matrix. Though the origin of ferromagnetism in these compounds is not totally understood, it was proposed that the exchange interaction between the $Mn^{2+}$ ions was responsible for the ferromagnetism in these compounds. Recently, in reference [67], surface vacancy induced ferromagnetism is obtained for a series of oxide materials including ZnO. In the present situation also, defects may play important roles in obtaining ferromagnetism. A cation vacancy near Fe, can promote $Fe^{2+}$ into $Fe^{3+}$ and also mediate the $Fe^{2+}$ - $Fe^{2+}$ exchange interaction. Since the TM doping percentage is slightly on the higher side towards cationic percolation threshold, $Fe^{2+}$ - $Fe^{3+}$ exchange, although being less in number in comparison to $Fe^{2+}$ - $Fe^{2+}$ interaction, may also be possible.

Another possible source of $Fe^{3+}$ can be the presence of minute segregated impurity phases like spinel ferrites, remaining undetected by XRD. In room-temperature, the well-known spinel ferrite $ZnFe_2O_4$ is paramagnetic (with an antiferromagnetic ordering below 10 K) in bulk sample with all the $Fe^{3+}$ in the octahedral interstitial site in a closed packed lattice of oxygen. On the other hand, nanocrystalline nonstoichiometric $ZnFe_2O_4$ can have $Fe^{3+}$ both in tetradehral and octahedral sites and can result into a ferrimagnet with high magnetic moments[33]. However, for the present case, the EPR and Mossbauer studies reveal some more details. The line position, signal shape and nature of evolution of signal shape with temperature is different from the ferrimagnetic ferrite phase [64,68]. Moreover, the EPR signal for $Fe^{3+}$ is of paramagnetic nature with g = 2 and there is a possibility that EPR spectra may not resolve the signals for $Fe^{3+}$ from the tetrahedral and octahedral sites [69], as can be obtained from NMR studies [33,70]. Also, the isomer shift obtained for paramagnetic doublet is matching neither with the tetrahedral nor with the ocatahedral $Fe^{3+}$ isomer shift values in nanocrystalline ferrimagnetic zinc ferrites [64,71]. In the Fe-doped ZnO system, Fe ions are present in tetrahedral site. If due to the presence of cation vacancy, the valence state changes to 3+, the crystal field no longer posseses tetrahehral symmetry and hence the isomer shift may not match with the value corresponding to tetrahedral sites. Also, in reference [48], the associated secondary phase was $Fe_3O_4$, the formation of which had reduced the value of magnetic moment of the sample [48]. The presence of $Fe_3O_4/Fe_2O_3$ in the present sample is already nullified by verification of EPR spectra of those respective compounds.

Hence, in the present studies, the ferromagnetism obtained may be defect-induced. In the following section, we have tried to understand the present experimental situations with the help of ab-initio electronic structure calculations.

# VII. Theoretical Investigation:

## Structure and Computational details:

Pure ZnO crystallizes in the wurtzite structure, a hexagonal analog of zincblende lattice, with a space group $P6_3mc$ (No. 186) with two formula unit per unit cell, where each Zn-atom is tetrahedrally co-ordinated with four other O-atoms. The lattice constants [72] are $a = b = 3.2495\ A^0$ and $c = 5.2069\ A^0$ and the atomic positions for Zn are (0, 0, 0) and (1/3, 2/3, 0.5) and for O are (0, 0, 0.3408) and (1/3, 2/3, 0.8408). The analysis of electronic structure and magnetic properties are carried out in the framework of the tight-binding linear Muffin-tin orbital method (TB-LMTO) in the atomic sphere approximation (ASA)[46] within LSDA [47,48]. The space filling in the ASA is achieved by inserting empty spheres at the interstitials and by inflating the atom-centered non-overlapping spheres. The atomic radii are chosen such that the charge on the empty spheres are negligible and the mutual overlap between all kinds of combinations of interstitial and atomic spheres are within the permissible limit of the ASA. In order to study the effects of TM doping in this system, we have constructed supercell with the size being dependent on the percentage of TM doping. In the present case, we have performed a (2 × 2 × 2) supercell calculation with 64 atoms (including empty spheres) and sixteen formula units of ZnO. To simulate a TM doped system, we have replaced the appropriate number of Zn atoms with TM atoms. The size of the supercell is chosen to achieve the doping percentage comparable to our experimental data. For all these self-consistent calculations, the k-point mesh size is (12, 12, 8). The calculations have also been carried out and verified with a smaller supercell with 32 atoms and with 8 formula unit of **ZnO**.

**Results and Discussions**

In this section, we shall discuss the electronic structure of Fe doped ZnO both with and without defects (*i.e.* Zn and O vacancies) in order to understand the origin of ferromagnetism in these systems. Although, our calculations are for the bulk system, however, it will provide important insights about the origin of intrinsic ferromagnetism in the Fe doped ZnO nanocrystals considered in this work.

**Fe doped ZnO**

In Fig. 13 we have displayed the paramagnetic DOS for 12.5% Fe doped ZnO without any defects. The characteristic feature of the DOS is the deep Fe derived states in the semiconducting gap of ZnO. These states are broadened due to hybridization with O-p states. The expected valence state of the substitutional Fe atom will be 2+ and it will be in the $d^6$ configuration. The Fe-d states in the tetrahedral co-ordination gets split into two fold degenerate $e$-levels and three fold degenerate $t_2$-levels with the $e$ levels being lower in energy. So, out of the available 6 electrons per Fe, four electrons are accommodated in the low lying $e$-level and the rest into the $t_2$-levels in the non-spin polarized calculation as can be seen in Fig. 13(a).

The resulting partially filled $t_2$-levels have appreciable density of states at Fermi energy, $D(E_F) \sim 12.84$ states / eV / Fe making the paramagnetic state unstable and susceptible to stoner instability. The stability, however, can be achieved by incorporating magnetic order into the system with spin polarization and thereby resulting in a difference between the population of the majority and minority spin channels due to electronic rearrangement. The results of the spin-polarized calculations are displayed in Fig. 13(b). We find the majority spin channel (spin-down) is fully occupied, while the minority spin channel is only partly occupied resulting in a half metallic ferromagnet with 100% spin polarization. We note from the figure, that the exchange splitting is much larger in comparison to the crystal field splitting and the system therefore favors a high spin configuration with the various levels serially filled up as $e{\downarrow}(2)$, $t_2{\downarrow}(3)$, $e{\uparrow}(1)$, (where the numbers in the parenthesis indicate the number of electrons). Hence, the spin down channel completely filled while the spin up $e{\uparrow}$ level is half filled resulting in a net integral magnetic moment of 4 $\mu_B$. However, such a situation cannot stabilize ferromagnetism in the presence of Coulomb correlations. If the Coulomb correlations are included, as in the LSDA+U method, the half filled $e{\uparrow}$ band will split and the system will be an insulator. In fact, a recent LSDA+U [73] calculation indicated an insulating antiferromagnetic state to be more stable in comparison to the ferromagnetic state. Hence, from the preceeding discussion we conclude that Fe doped ZnO is unlikely to stabilize in a ferromagnetic state. In view of our experimental results, as discussed in the previous section, supporting the presence of defects in the system, it is very likely that the defects plays a crucial role to stabilize ferromagnetism in this system. In the following section we have, therefore, analyzed the electronic structure of Fe doped ZnO in the presence of the defects, *viz*. (i) O vacancy, (ii) Zn vacancy.

**Fe Doped ZnO with defects**

Figure 14 displays the total as well as the partial DOS for 12.5% Fe doped ZnO with 12.5% O vacancy (top, (a) and (b) ) and 12.5% Zn vacancy (bottom, (c) and (d)) with single Fe and single vacancy per supercell. It is evident from the Fig. 14(a) and (b), that the oxygen vacancies produce shallow donor states while the zinc vacancies produce shallow acceptor states. These states are delocalized due to the hybridization with the Fe-d states. An oxygen vacancy adds electrons to the system and thereby effectively dopes n-type carriers to the system. These electrons are accommodated in the minority spin channel (spin up), so in comparison to the Fe doped ZnO, the minority spin channel has fully occupied $e\uparrow$ level and singly occupied $t_2\uparrow$ level as can be seen in Fig. 14(b). On the other hand Zn vacancies dope hole to the system, resulting in a completely empty minority spin channel (spin-up) and one hole in the majority spin channel (spin-up) (see Fig 14(d)). Such a situation is condusive for ferromagnetism mediated by double exchange as argued by Akai, [7] where, if a d-orbital is partly occupied then the electrons in that orbital are allowed to hop to the neighboring d-orbitals, provided the neighboring Fe atoms are in parallel spin-configuration. Thus the d-electrons lower the kinetic energy by hopping in the ferromagnetic state. On the other hand, if the d-orbital is completely occupied then this reduction of energy via hopping is not possible and the system energy gets lowered by antiferromagnetic spin alignment of the neighboring Fe atoms by super-exchange [74]. The system can appreciably lower its energy by double exchange, when it is near the half filling with sufficient (usually small) number of holes (electrons). So we expect that hole doping by Zn vacancies will be more effective to stabilize ferromagnetism in this system.

In order to explore the possibilities of ferromagnetic versus anti-ferromagnetic ordering in Fe doped ZnO in the presence of O or Zn vacancies, we have performed spin-polarized density functional calculations with (i) two Fe spins parallel to each other, the ferromagnetic configuration, (ii) two Fe spins anti-parallel to each other, the anti-ferromagnetic configuration in the presence of O and Zn vacancies. We have also calculated the energy difference $\Delta E$ as a function of Fe-Fe separation. This energy difference $\Delta E$ is also a measure of the inter-atomic exchange interaction and in the framework of mean-field theory is also proportional to $T_c$ for the system. The results of our calculation are displayed in Fig. 16. We note both for the O vacancy as well as the Zn vacancy, the nearest neighbor Fe-Fe interaction is ferromagnetic, however, this exchange coupling is much stronger for the hole doped system with Zn vacancy, according to our expectation. The further neighbor interactions are antiferromagnetic for the electron doped system (with oxygen vacancy), suggesting lower probability for electron doped system to be a ferromagnet in agreement with the recent LDA + U calculations [73]. On the other hand, the further neighbour interactions for the hole doped system with Zn vacancy is ferromagnetic but are much weaker in comparison to the nearest neighbor interaction. Such short- range exchange interaction indicates the formation of Fe clusters and possibility of intrinsic ferromagnetism in Fe doped ZnO to be driven by percolation.

In conclusion, our calculations indicate that hole doping, possibly by Zn vacancies, is crucial to stabilize ferromagnetism in Fe doped ZnO. As we have argued in the experimental section also, for Fe doped ZnO nano- crystals, cation vacancies may be present and thereby effectively dope hole into the system. A possible signature of hole doping may be the presence of $Fe^{3+}$ in our samples as confimed by EPR measurements and Mössbauer spectroscopy.

# VIII. Conclusions

We have successfully synthesized Fe-doped ZnO nanocrystals and structurally characterize it by XRD and TEM. The magnetic measurements show the presence of room temperature ferromagnetic order within the system. The presence of spin-glass behaviour at low temperature is explained by a core-shell spin structure of the individual underlying nanoparticle system. We have tried to understand the exact origin of ferromagnetism with the help of local probes like EPR and Mössbauer, which indicates the presence of the dopant cation in both valence states 2+ and 3+. The ab-initio electronic structure calculations for the same material suggest that hole doping is crucial to promote ferromagnetism in this system and the presence of $Fe^{3+}$ in our samples is a possible signature of hole- doping induced by Zn vacancies.

**Acknowledgements:**

DK is grateful to Dr. V. C. Sahni, Dr. J. V. Yakhmi, Dr. S. K. Gupta and Dr. V. K. Manchanda for their encouragement and support for this work. ID thanks DST. India (project SR/S2/CMP-19/2004) for financial support. AKD thanks Dr. B. Satpati and Dr.

P. V. Satyam for TEM measurements. GPD would like to thank Prof. K. V. Rao for his comments and helpful discussions.

**References:**

[1] S. A. Wolf, D. D. Awschalom, R. A. Buhrman, J. M. Daughton, S. Von. Molnar, M. L. Roukes, A. Y. Chtchelkanova, D. M. Treger;  Science, **294**, 1488 (2001).

[2] H. Ohno et. al., Appl. Phys. Lett. **69**, 363 (1996).

[3] H. Ohno et. al., Phys. Rev. B **57**, 2037 (1998).

[4] Hayashi et.al., Appl. Phys. Lett., **78**, 1691 (2001).

[5] Potashnik et.al., Phys. Rev. B **66**, 012408 (2002).

[6] T. Dietl, H. Ohno, F. Matsukura, J. Cibert and D. Ferrand, Science **287**, 1019 (2000).

[7] H. Akai, Phys. Rev. Lett. **81**, 3002 (1998).

[8] Mark van Schilfgaarde and O. N. Mryasov Phys. Rev. B **63**, 233205 (2001).

[9] J. K. Furdyna, J. Appl. Phys., **64**, R29 (1988).

[10] K. Ueda, H. Tabata and T. Kawai, Appl. Phys. Lett. **79**, 988 (2001).

[11] S. B. Ogale et. al., Phys. Rev. Lett., **91**, 077205 (2003).

[12] Y. Matsumoto et. al., Science **291**, 854 (2001).

[13] C. Song, K. W. Geng, F. Zeng, X. B. Wang, Y. X. Shen, F. Pan, Y. N. Xie, T. Liu, H. T. Zhou and Z. Fan, Phys. Rev. B **73**, 024405 (2006).

[14] N. H. Hong, J. Sakai, N. T. Huyong, N. Poirot and A. Ruyter, Phys. Rev. B **72**, 045336 (2005).

[15] M. A. Garcia, M. L. Ruiz-Gonzalez, A. Quesada, J. L. Costa-Kramer, J. F. Fernandez, S. J. Khatib, A. Wennberg, A. C. Cabellero, M. S. Martin-Gonzalez, M. Villegas, F. Briones, J. M. Gonzalez-Calbet and A. Hernando, Phys. Rev. Lett. **94**, 217206 (2005).

[16] K. R. Kittilstved, N. S. Norberg and D. R. Gamelin, Phys. Rev. Lett. **94**, 147209 (2005).

[17] M. Venkatesan, C. B. Fitzerald, J. G. Lunney and J. M. D. Coey, Phys. Rev. Lett. **93**, 177206 (2004).

[18] Q. Wang, Q. Sun, P. Jena and Y. Kawazoe, Phys. Rev. B **70**, 052408 (2004).

[19] A. S. Risbud, N. A. Spaldin, Z. Q. Chen, S. Stemmer and Ram Seshadri, Phys. Rev. B **68**, 205202 (2003).

[20] S. Kolenisk, B. Dabrowski and J. Mais, J. Appl. Phys. **95**, 2582 (2004).

[21] S. Kolenisk and B. Dabrowski, J. Appl. Phys. **96**, 5379 (2004).

[22] G. Lawes, A. S. Risbud, A. P. Ramirez and R. Seshadri, Phys. Rev. B **71**, 045201 (2005).

 [23] S. Yin, M. X. Xu, L. Yang, J. F. Liu, H. Rosner, H. Hann, H. Gleiter, D. Schild, S. Doyle, T. Liu, T. D. Hu, E. Takayama-Muromachi and J. Z. Jiang, Phys. Rev. B **73**, 224408 (2006).

[24] Kazunori Sato and Hiroshi Katayama-Yoshida, Jpn. J. Appl. Phys, **39** L555 (2000).

[25] Kazunori Sato and Hiroshi Katayama-Yoshida, Jpn. J. Appl. Phys, **40** L334 (2001).

[26] Tetsuya Yamamoto and Hiroshi Katayama-Yoshida, Jpn. J. Appl. Phys, **38** L166 (1999).

[27] V. Jayaram, J. Rajkumar and B. Sirisha Rani, J. Am. Ceram. Soc. **82**, 473 (1999).

[28] T. Fukumura, Z. Jin, A. Ohtomo, H. Koinuma and M. Kawasaki, Appl. Phys. Lett. **75**, 3366(1999).

[29] P. Sharma, A. Gupta, K. V. Rao, F. J. Owens, R. Sharma, R. Ahuja, J. M. O. Guillen, B. Johansson and G. A. Gehring, Nat. Mater. **2**, 673 (2003).

[30] P. V. Radovanic and D. R. Gamelin, Phys. Rev. Lett. **91**, 157202 (2003).

[31] T. Fukumura, H. Toyosaki and Y. Yamada, Semicond. Sci. Technol. **20**, S103 (2005).

[32] I. S. Elfimov, S. Yunoki and G. A. Sawatzky, Phys. Rev. Lett. **89**, 216403 (2002).

[33] Jeong Hyun Shim *et. al.* , Appl. Phys. Letts. **86**, 082503 (2005).

[34] S. Kolenisk, B. Dabrowski and J. Mais, Journal of Superconductivity: Incorporating Novel Magnetism **15**, 251 (2002).

[35] J. M. D. Coey, M. Venkatesan and C. B. Fitzerald, Nat. Mater. **4**, 173 (2005).

[36] A. Kaminski and S. Das Sarma, Phys. Rev. Lett. **88**, 247202 (2002).

[37] L. Bergqvist, O. Erikson, J. Kudrnovsky, V. Drchal, P. Korzhavyi and I. Turek, Phys. Rev. Lett. **93**, 137202 (2004).

[38] Marcel H. F. Sluiter, Y. Kawazoe, Parmanand Sharma, A. Inoue, A. R. Raju, C. Rout and U. V. Waghmare, Phys. Rev. Lett. **94**, 187204 (2005).

[39] S. W. Yoon, S. B. Cho, S. C. We, S.Yoon and B. J. Suh, J. Appl. Phys. **93**, 7879 (2003).

[40] S-J. Han, J. W. Song, C-H. Yang, S. H. Park and Y. H. Jeong, Appl. Phys. Lett. **81**, 4212 (2002).

[41] M. S. Park and B. I. Min, Phys. Rev. B **68**, 224436 (2003).

[42] P. Sati, R. Hayn, R. Kuzian, S. Schafer, A. Stepanov, C. Morhain, C. Deparis, M. Laugt, M. Gorian and Z. Golacki, Phys. Rev. Lett., **96**, 017203 (2006).

[43] B. Martinez, F. Sandiumenge, Ll. Balcells, J. Arbiol, F. Sibieude and C. Monty, Phys. Rev. B **72**, 165202 (2003).

[44] O. D. Jayakumar, H. G. Salunke, R. M. Kadam, Manoj Mahapatra, G. Yashwant and S. K. Kulshreshtha, Nanotechnology **17**, 1278 (2006).

[45] R. H. Kodama and A. E. Berkowitz, Phys. Rev. **B 59**, 6321 (1999).

[46] O. K. Andersen, O. Jepsen, Phys. Rev. Lett. **53,** 2571 (1984).

[47] U. von Barth, J. Phys. C:Solid State. Phys. **5**, 1629(1972).

[48] L. Hedin, J. Phys. C:Solid State. Phys. **4**, 2064(1971).

[49] S. K. Mandal, A. K. Das, T. K. Nath and Debjani Karmakar, Appl. Phys. Letts. **89**, 144105 (2006).

[50] A. C. Larson, R. B. Von Dreele, GSAS: general structure analysis system, Los Alamos National Laboratory, LAUR publication, 1998.

[51] L. Neel, J. Phys. Radium **12**, 225(1954).

[52] S. Das Sarma, E. H. Hwang and A. Kaminski, Phys. Rev. B **67**, 155201(2003).

[53] S. R. Shinde, S. B. Ogale, J. S. Higgins, H. Zheng, A. J. Millis, V. N. Kulkarni, R. Ramesh, R.L.Greene and T.Venkatesan, Phys. Rev. Lett. **92**, 166601 (2004).

[54] X.-G. LI, X. J. Fan and G. Ji, W. B. Wu, K. H. Wong, C. L. Choy and H. C. Ku, J. Appl. Phys., **85**, 1663 (1999).

[55] Diandra L. Lesile-Pelecky and Reuben D. Rieke, Chem. Matter. **8**, 1770 (1996) and references therein.

[56]  C. A. Cardoso, F. M. Araujo-Moreira, V. P. S. Awana, E. Takayama-Muromachi, O. F. de Lima, H. Yamauchi and M. Karppinen, Phys. Rev. B **67**, 020407(R) (2003).

[57] T. Zhu, B. G. Shen, J. R. Sun, H. W. Zhao and W. S. Zhan, Appl. Phys. Lett. **78**, 3863 (2001).

[58] F.J.Owens, J. Phys. Chem. of Solids **66**, 793 (2005).

[59]  R.M.Kadam, M.K.Bhide, M.D.Sastry, J.V.Yakhmi, O.Kahn; Chem. Phys. Lett., 357, 457 (2002).

[60] N.S.Norberg, K.R.Kittilstved, J.E.Amonette, R.K.Kukkadapu, D.A.Schwartz  and D.R. Gamelin,  J. Am. Chem.  Soc. **126**,  9387 (2004).

[61]  U. Kaufmann, Phys. Rev. B  **14**,  1848 (1976).

[62]  Shekhar D. Bhame, V.L.Joseph Joly and P.A.Joy;  Phys. Rev. B **72**, 054426 (2005).

[63] G. F. Goya and E. R. Leite, J. Phys. Condens. Matter **15**,  641 (2003).

[64] R. N. Bhowmik, R. Ranganathan, R. Nagarajan, Bishwatosh Ghosh and S. Kumar, Phys. Rev. B **72**, 094405 (2005).

[65] K. Potzger, Shengqiang Zhou, H. Reuther, A. Mucklich, F. Eichhorn, N. Schell, W. Skorupa, M. Helm, J. Fassbender, T. Herrmannsdorfer and T. P. Papageorgiou, Appl. Phys. Letts. 88, 052508 (2006).

[66] C. B. Fitzerald, M. Venkatesan, A. P. Douvalis, S. Huber,  J. M. D. Coey and T. Bakas, J. Appl. Phys. **95**, 7390 (2004).

[67] A. Sundaresan *et. al.* Phys. Rev. **B 74**, 161306R (2006).

[68] J. F. Hochepied, M. P. Pileni, J. Magn. Magn. Mater., **231** 45 (2001).

[69] C. B. Azzoni *et. al.* Solid State Commn. **117** 511 (2001).

[70] Jeong Hyun Shim *et. al.*, Phys. Rev. **B 73,** 064404 (2006).

[71] C. N. Chinnasamy *et. al.*, J. Phys.: Condens. Matter **12**, 7795 (2000).

[72] R. W. G. Wyckoff: Crystal Structures (Wiley, New York, 1986) 2$^{nd}$ ed., Vol. 1, p.112.

[73] Priya Gopal and Nicola A. Spaldin, arXiv: cond-mat/ 0605543 v1 22 May (2006).

[74] P. W. Anderson and H. Hasegawa, Phys. Rev. **100**, 675 (1955).

**Figure Captions:**

**Fig. 1**: (a) The experimental (×) and Le-Bail fitted XRD data (solid line) for pure ZnO nanocrystalline sample is plotted. (b) Same graph for 10 % Fe- doped ZnO. For each cases, the difference plots are shown below.

**Fig. 2**: Using the cell parameter refinement with Le-Bail fitting, the evolution of the cell volume is plotted as a function of Fe-concentration to indicate the solubility of Fe in the ZnO matrix.

**Fig. 3**: (a) Low resolution TEM micrograph of $Zn_{0.9}Fe_{0.1}O$ nanoparticles calcined at a temperature of $350^{0}C$. (b) SAD patterns of the same sample showing single crystalline spots. (c) High resolution TEM micrograph for the same showing the single crystalline nature of each nanoparticle. (d) Nanometric particle size distribution of the same sample fitted with a lognormal distribution function, from which the estimated average particle size is ~ 7 nm.

**Fig. 4**: The magnetization –vs- temperature curves for $Zn_{0.9}Fe_{0.1}O$ in the field-cooled (FC) and zero-field cooled (ZFC) condition for the applied field value (a) 100 Oe, (b) 1000 Oe, (c) 1500 Oe, (d) 5000 Oe. Graph 4(a) indicates that the Ferro-to- para transition temperature is well above room temperature. For the first three graphs, the system undergoes a spin-glass transition at low temperature regime.

**Fig. 5**. (a) Relaxation of thermoremnant magnetization is shown for the sample after field cooling it at a field value 1000 Oe to confirm the spin-glass behaviour at low temperature. (b) The spin-glass arrangement of the spins due to surface spin disorder and the field or temperature induced crossover from spin-glass to ferromagnetic state is schematically explained by means of the simple core-shell model, which is diagrammatically presented in this figure.

**Fig. 6.** (a) Inverse dc susceptibility ($1/\chi_{dc}$) is plotted as a function of temperature for two applied field values 1000 Oe and 1500 Oe. The straight lines represent the Curie-Weiss fit to the same plots. (b) Inverse Differential susceptibility (corresponding to the fields 1000 Oe and 1500 Oe) is plotted as a function of temperature. The dotted line represents the fit with the paramagnetic Curie behaviour and the dashed line represents a Curie-Weiss fit.

**Fig. 7.** (a) The Magnetization versus field hysteresis loops are plotted for 5K, 270 K and 300 K for a comparison of the low and high temperature loops. For low temperature loops the paramagnetic contribution is high.

(b) The low temperature hysteresis loops are plotted at 2K, 5K and 20 K.

(c) The figure depicts the high field loop behaviour ( ± 12 T) at 2 K, where there is an approach towards saturation.

**Fig. 8**. The experimental and fitted EPR spectra are depicted at (a) room temperature and at (b) 100 K. The spectra is seen to be composed of two signals, viz., one intense and broad Gaussian signal (signal *A*) and an weak and narrow Lorentzian signal (signal *B*). The inset represents a measure of the number of spins participating in producing signal *A* and signal *B*.

**Fig. 9**. Evolution of EPR spectra as a function of temperature from 100 K to 450 K is shown in the figure. EPR line intensity is plotted as the y-axis. With increase of temperature the line-shape becomes more symmetric and resonance field increases to higher fields.

**Fig. 10a**. The (a) line width ($\Delta H_{pp}$), (b) peak area (in arbitrary units), (c) intensity (in arbitrary units) and (d) resonance field or line position ($H_r$) are plotted as a function of temperature for signal *A*.

**Fig. 10b**. The (a) peak area and (b) intensity are plotted in arbitrary units as a function of temperature for signal $B$.

**Fig. 11**. The evolution of Mossbauer spectra as a function of temperature is depicted in figure recorded at three different temperatures (a) room temperature, (b) 12 K and (c) 4K.

**Fig. 12**. The Mossbauer spectra recorded at 12 K is shown in a more elaborated fashion where the positions of the hyperfine sextet splitting are shown. Also the quadrupole splittings of both $Fe^{2+}$ and $Fe^{3+}$ are indicated alongwith the paramagnetic doublet.

**Fig. 13**. The (a) non-spin polarized and (b) spin-polarized DOS's are plotted for Fe doped ZnO system. Within the framework of LSDA, the ferromagnetic calculation results into 100% spin polarization and partially filled majority spin levels.

**Fig. 14.** Figures depict a comparison of the total DOS and site projected DOS for the pure system and in the presence of O and Zn vacancy for single Fe and single vacancy per supercell. (a) total DOS for Fe-doped ZnO system in presence of O-vacancy with 12.5 % Fe and 12.5 % O-vacancies per supercell, (b) partial O-vacancy DOS (filled curve) and Fe-d projected DOS (solid line), (c) total DOS for Fe-doped ZnO system in presence of Zn-vacancy with 12.5 % Fe and 12.5 % Zn-vacancies (d) partial Zn-vacancy DOS (filled curve) and Fe-d projected DOS (solid line). The respective percentages are mentioned in figure.

**Fig. 15**: The energy difference between the AFM and FM ground states, which gives a measure for the exchange parameter J of the interaction between two Fe spins, is plotted as a function of the nearest neighbour distance between the two Fe-spins in presence of O-vacancy and Zn-vacancy.

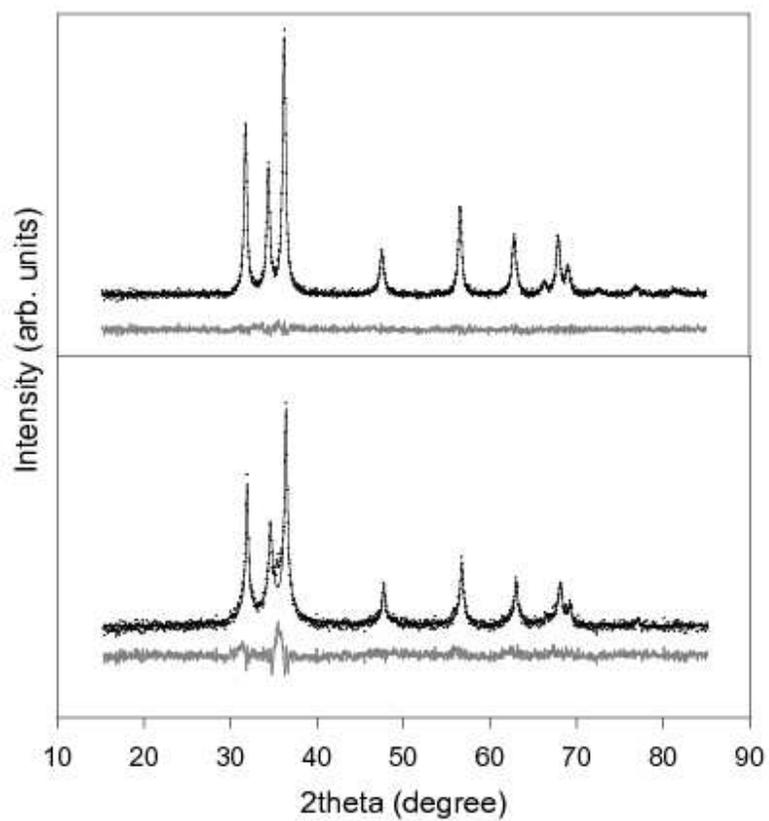

**Figure 1**

**Debjani Karmakar et al.**

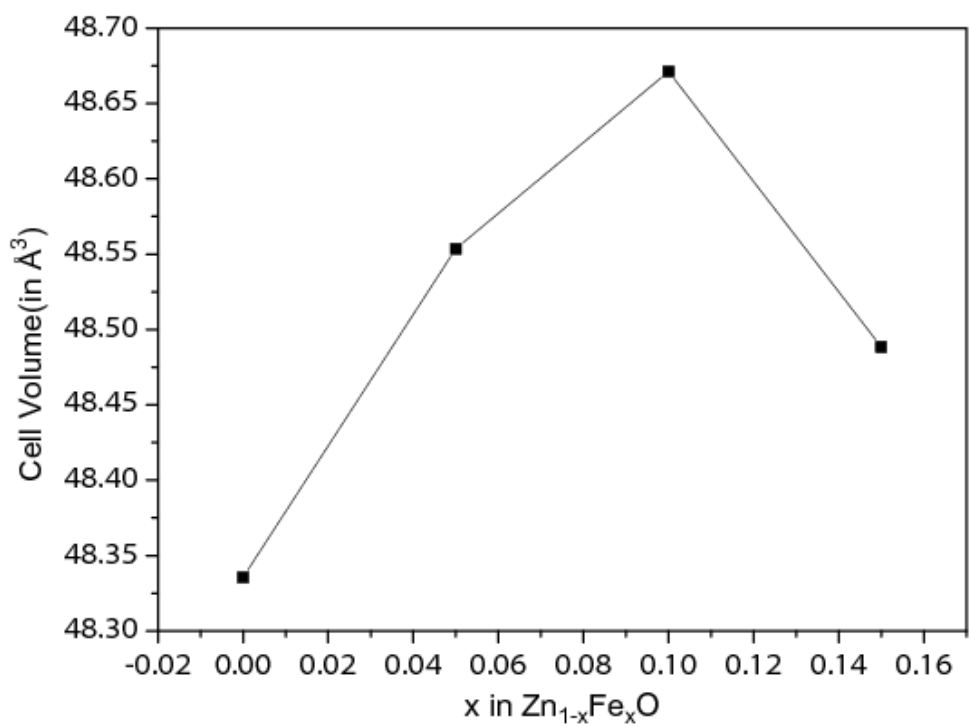

**Figure 2**

**Debjani Karmakar et al.**

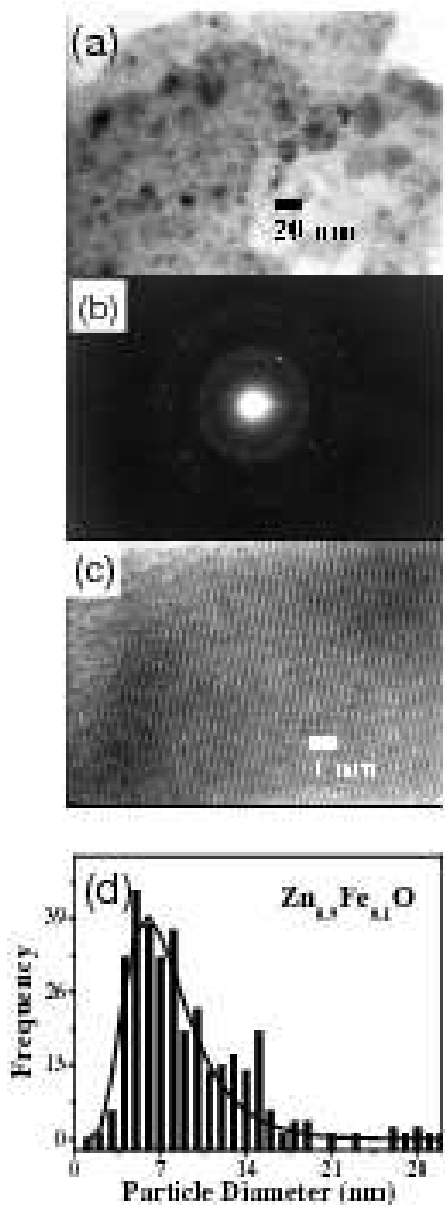

**Figure 3**

Debjani Karmakar et al.

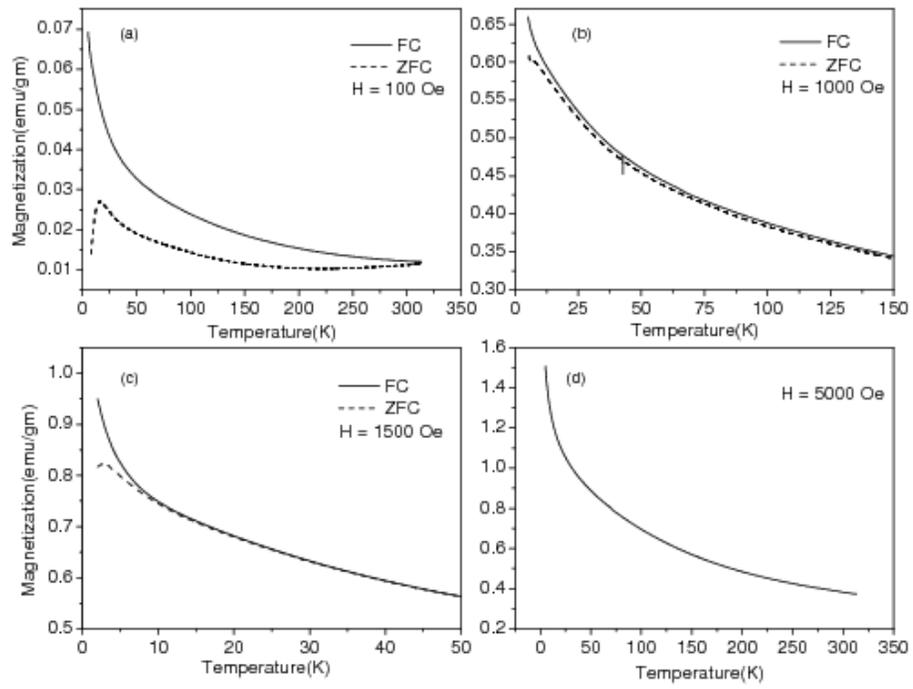

**Figure 4**

**Debjani Karmakar et al.**

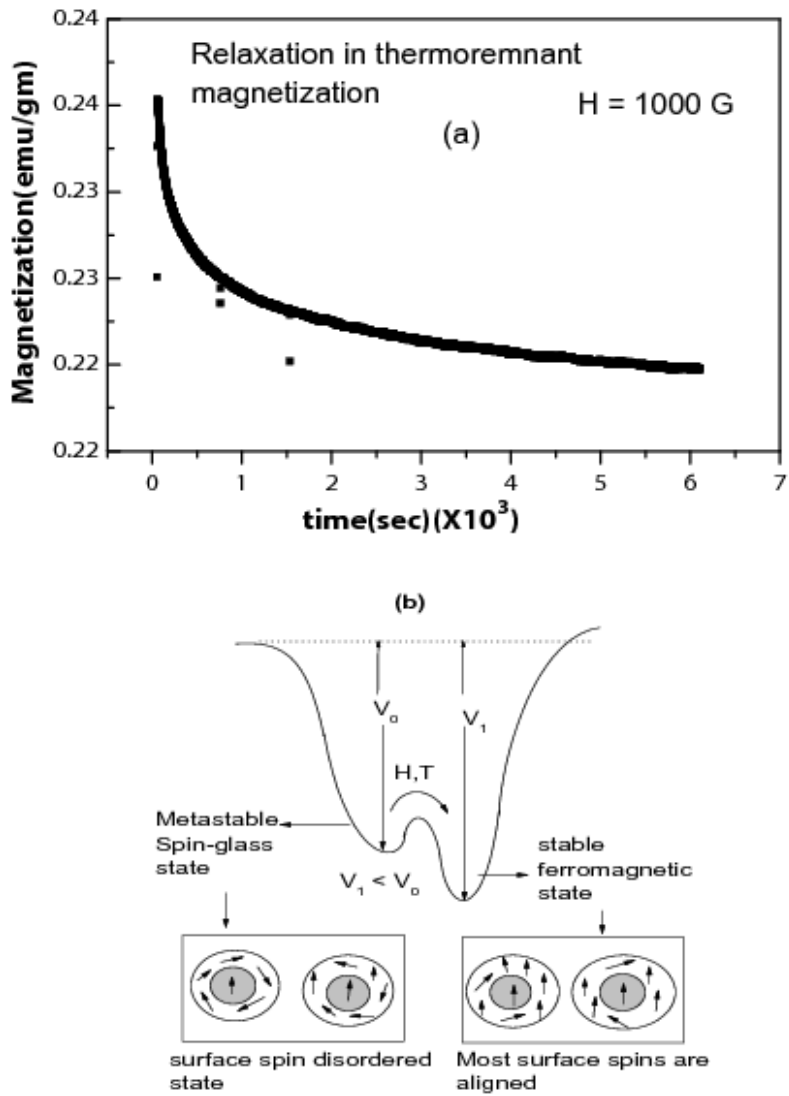

**Figure 5**

Debjani Karmakar et al.

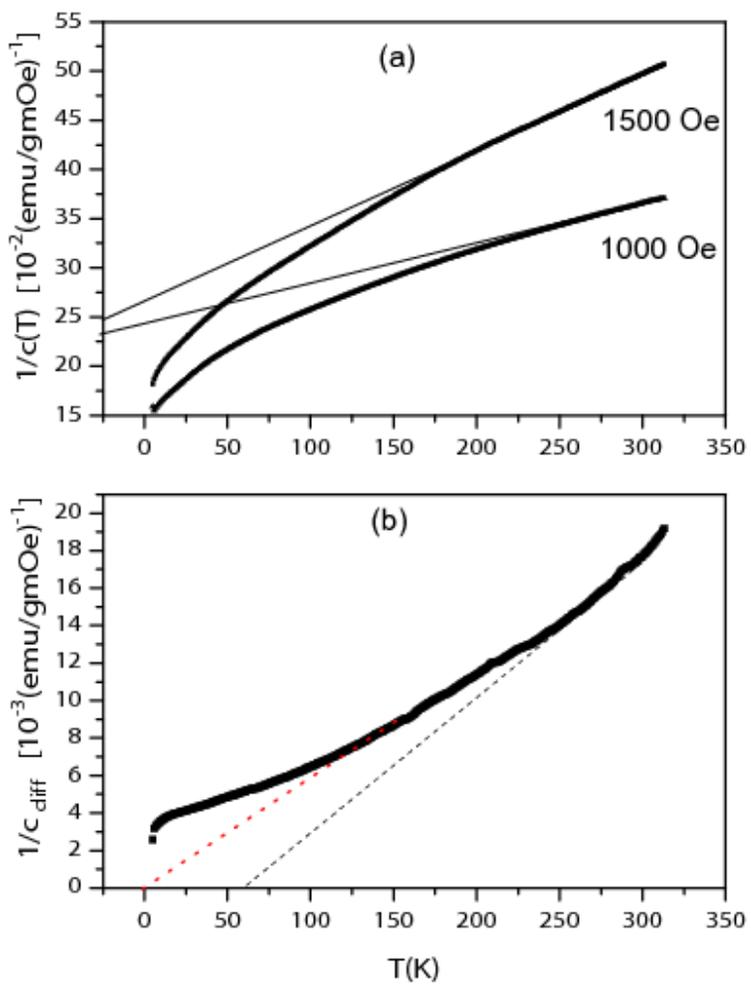

**Figure 6**

**Debjani Karmakar et al.**

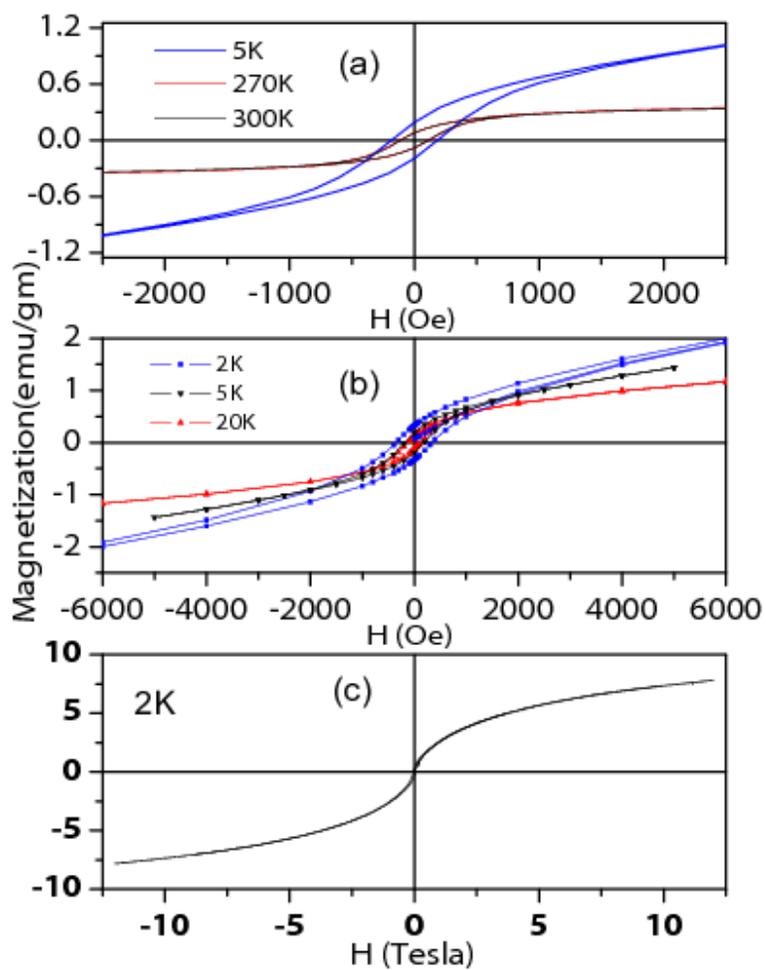

**Figure 7**

**Debjani Karmakar et al.**

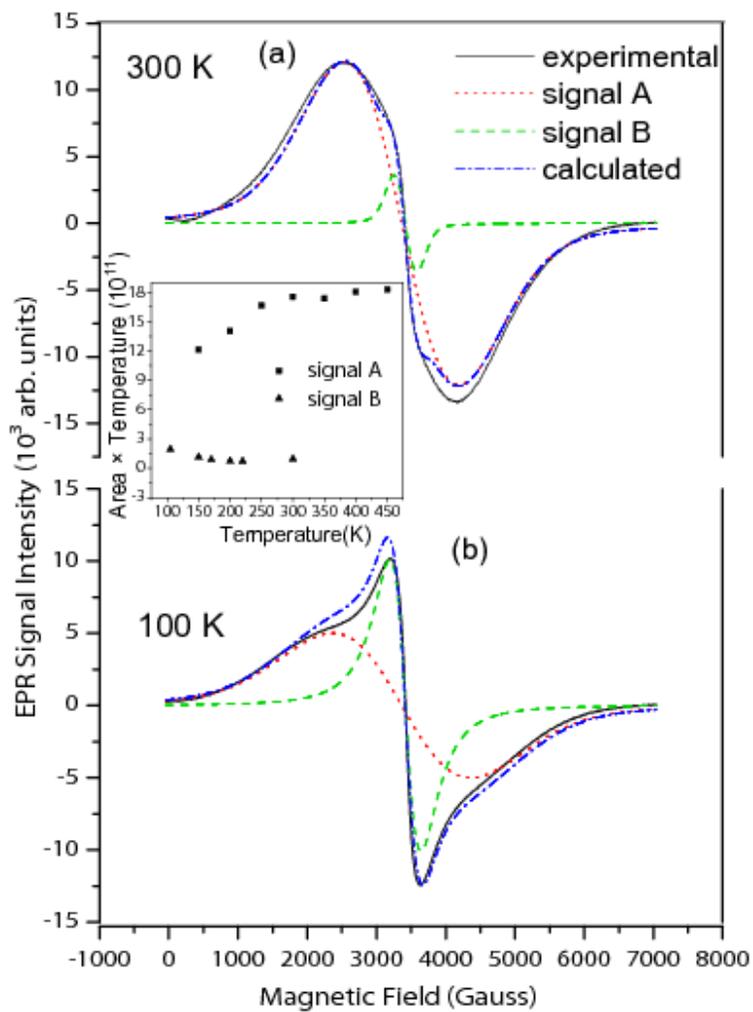

**Figure 8**

**Debjani Karmakar et al.**

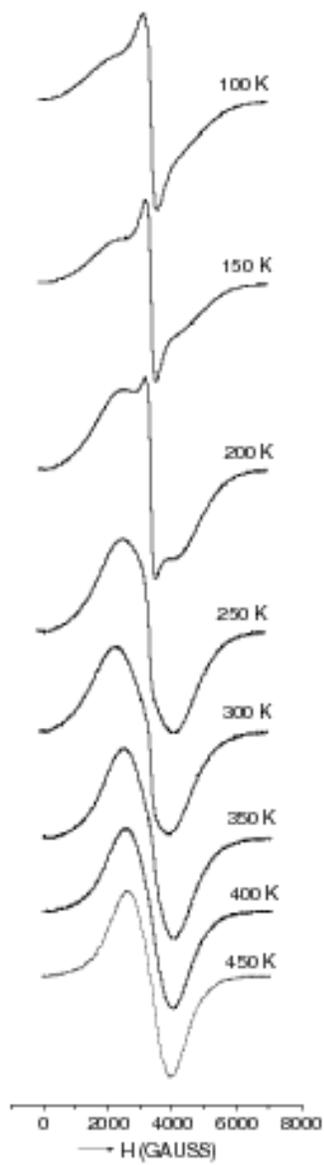

**Figure 9**

**Debjani Karmakar et al.**

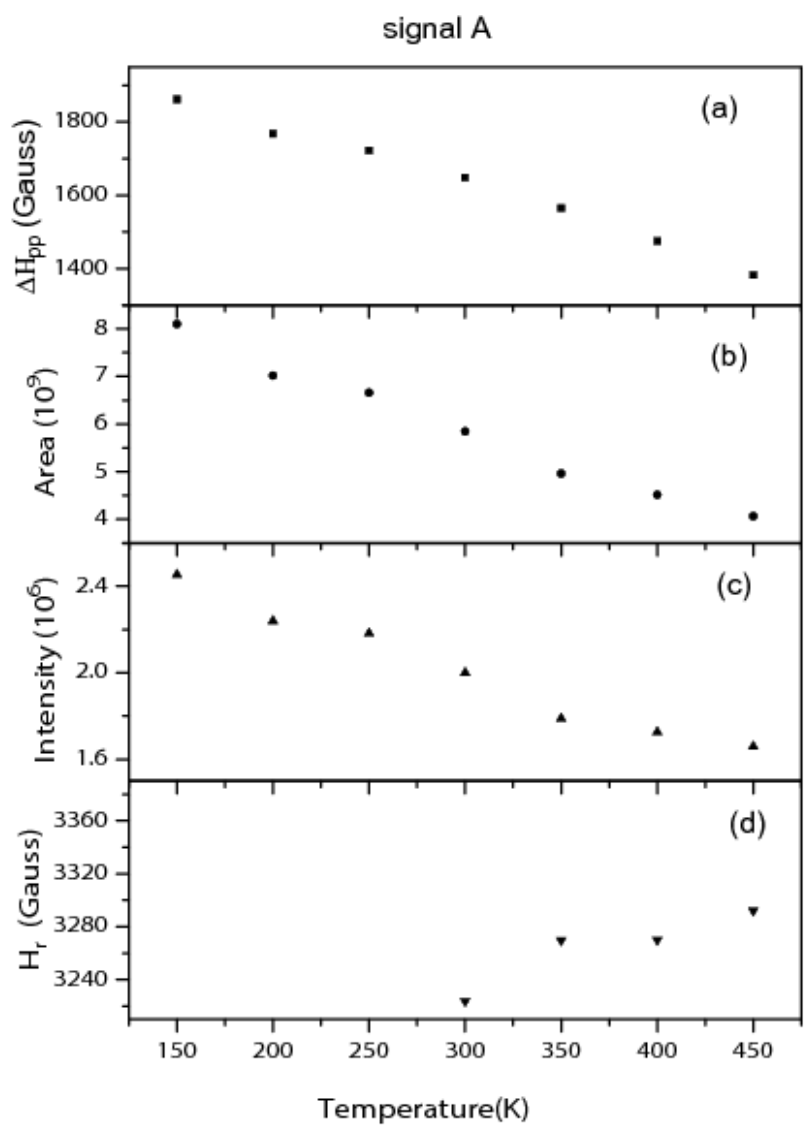

**Figure 10a**

**Debjani Karmakar et al.**

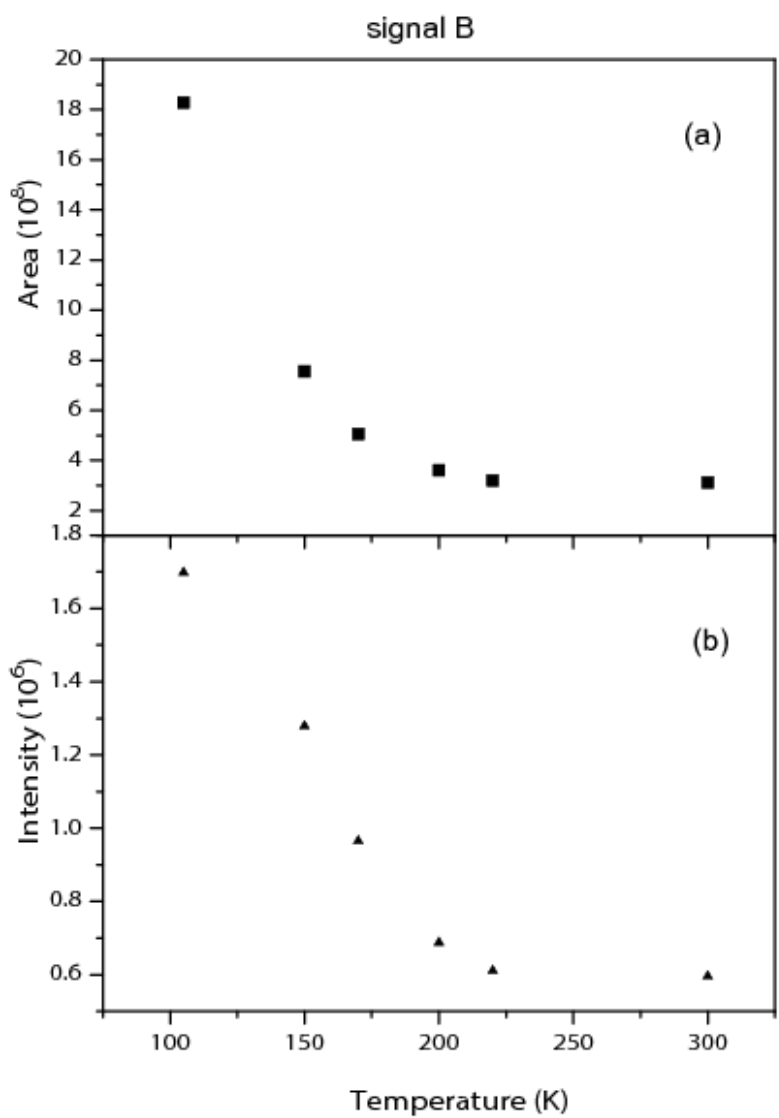

**Figure 10b**

**Debjani Karmakar et al.**

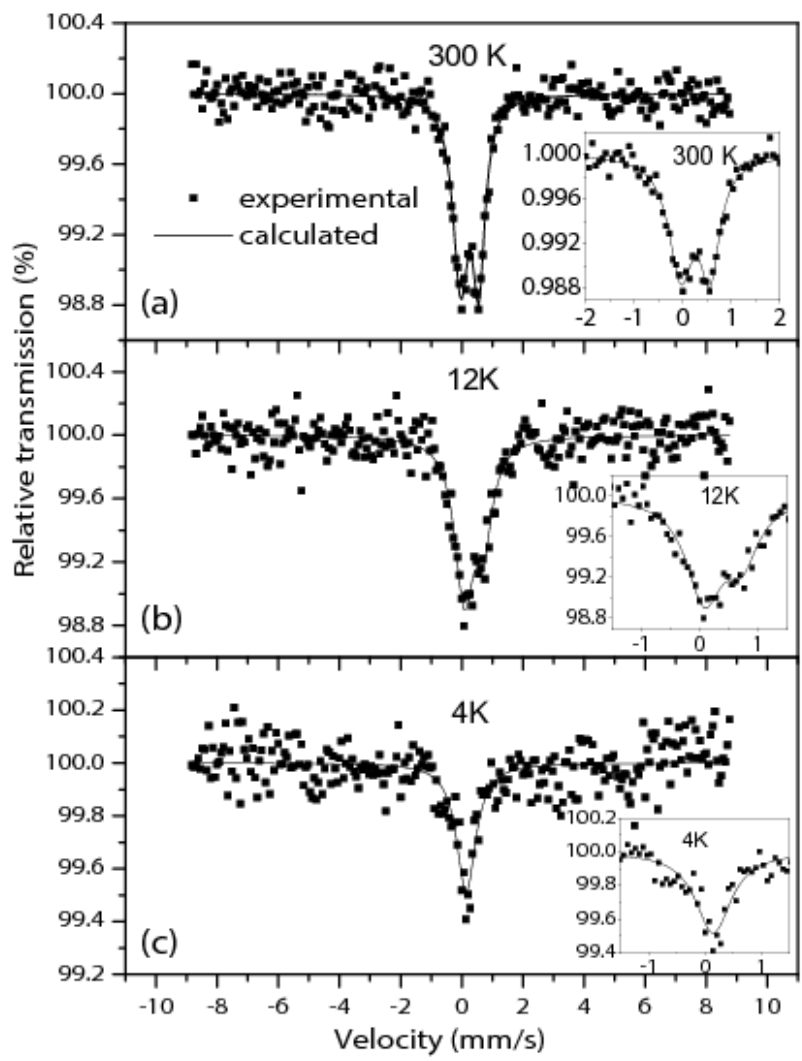

**Figure 11**

Debjani Karmakar et al.

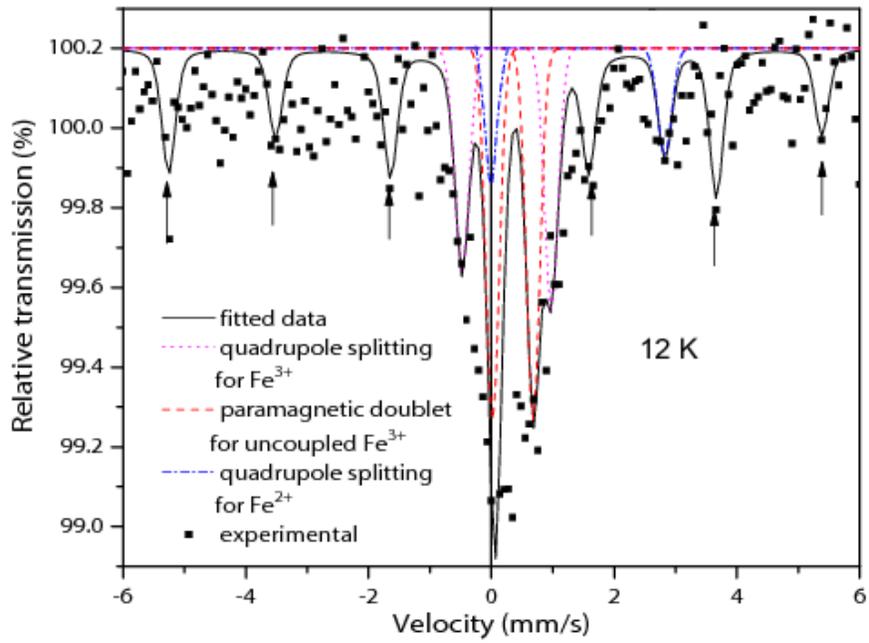

**Figure 12**

**Debjani Karmakar et al.**

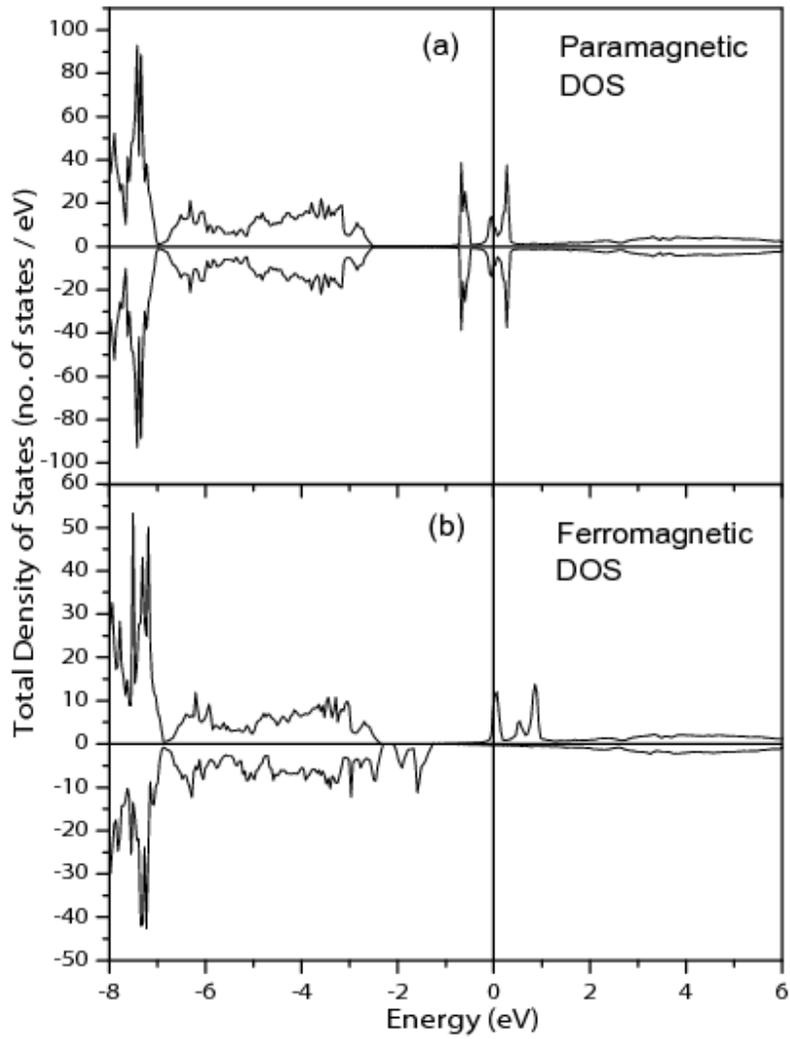

**Figure 13**

**Debjani Karmakar et al.**

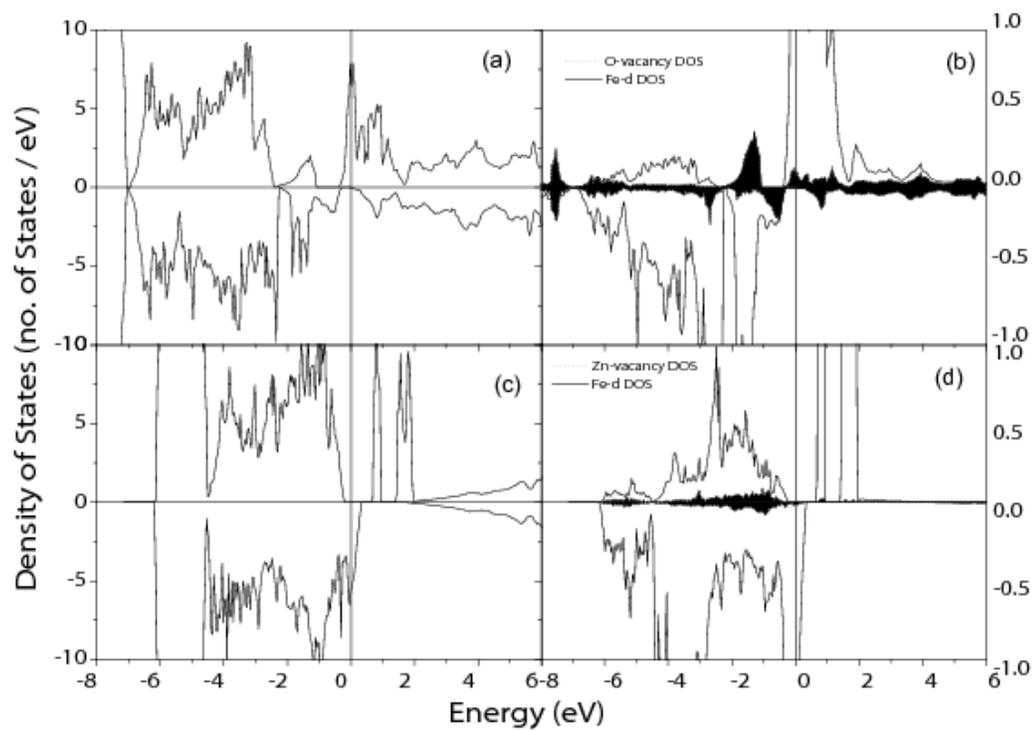

**Figure 14**

**Debjani Karmakar et al.**

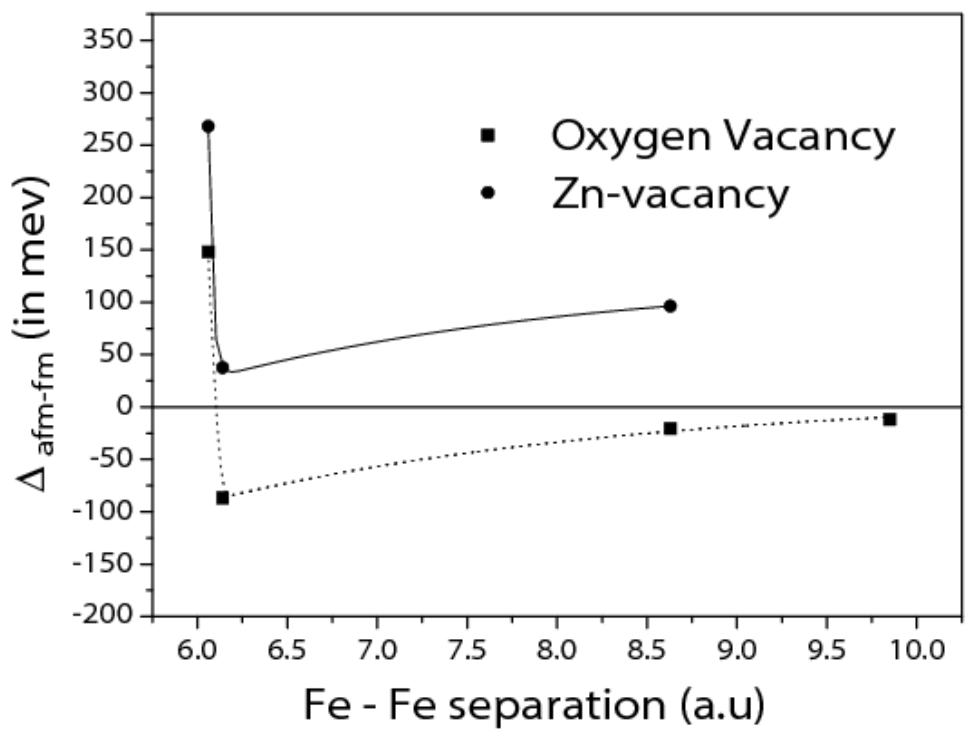

**Figure 15**

**Debjani Karmakar et al.**